\documentclass[aps,prl,amsmath,amssymb,aps, floatfix,reprint]{revtex4-1}
\usepackage{CJK}
\usepackage{graphicx}
\usepackage{dcolumn}
\usepackage{float}
\usepackage{fixltx2e}
\usepackage{bm}

\begin{document}
\begin{CJK*}{}{} 
\title{The Evolution of Flow and Mass Transport in 3D Confined Cavities}
\author{Reem Khojah}
\affiliation{ Bioengineering Department, University of California Los Angeles, Los Angeles, California 90095, USA
}%
\author{Darren Lo}
\affiliation{ Bioengineering Department, University of California Los Angeles, Los Angeles, California 90095, USA
}%
\author{Fiona Tang}
\affiliation{ Bioengineering Department, University of California Los Angeles, Los Angeles, California 90095, USA
}%
\author{Dino Di Carlo}
 \email{dicarlo@ucla.edu}

\affiliation{ Bioengineering Department, University of California Los Angeles, Los Angeles, California 90095, USA
}%

\date{\today}
\maketitle
\end{CJK*} 


\begin{abstract}

Flow in channels and ducts with adjoining cavities are common in natural and engineered systems. Here we report numerical and experimental results of 3D confined cavity flow, identifying critical conditions in the recirculating flow formation and mass transport over a range of channel flow properties ($0.1\leq Re \leq 300$) and cavity aspect ratio ($0.1 \leq H/X_{s} \leq1$). In contrast to 2D systems, a mass flux boundary is not formed in 3D confined cavity-channel flow. Streamlines directly enter a recirculating vortex in the cavity and exit to the main channel leading to an exponential increase in the cavity mass flux when the recirculating vortex fills the cavity volume. These findings extend our understanding of flow entry and exit in cavities and suggest conditions where convective mass transport into and out of cavities would be amplified and vortex particle capture reduced.

\end{abstract}
\pacs{Valid PACS appear here}
\maketitle

When flowing fluid with finite inertia in a channel that suddenly expands in cross-sectional dimension, an adverse pressure gradient can occur resulting in the formation of recirculating flow in the expansion region \cite{batchelor2000introduction,macagno1967computational, baloch1995two}.  If the cross-sectional dimension is returned to the initial value further downstream, this creates a cavity, which can support a recirculating flow up to the cavity size \cite{sinha1982laminar}. The formation of recirculating flows in 3D-confined cavities can result in selective particle capture from the mainstream channel flow, an area attracting significant interest in biotechnology and biomedicine where microcavities have been utilized as well-controlled reaction chambers for trapped bio-particles in miniaturized diagnostic devices \cite{marcus2006parallel, gierahn2017seq, hur2011high, khojah2017size, dhar2018functional}. Moreover, cavity flow capture is a diverse phenomenon which can also be found in nature such as in the assembly of biological self-replicating molecules inside microvortices that arise in porous ocean rocks \cite{sun2018control}, as well as platelet aggregates accumulation in aneurysms \cite{rayz2008numerical} (Figs 1(a-c)).  Thus, a better understanding of recirculating cavity flow can elucidate trapping phenomena in natural and physiological flows and enable the engineering of mass transport in cavities and microwells.

\begin{figure}[t]
\includegraphics[scale=0.08]{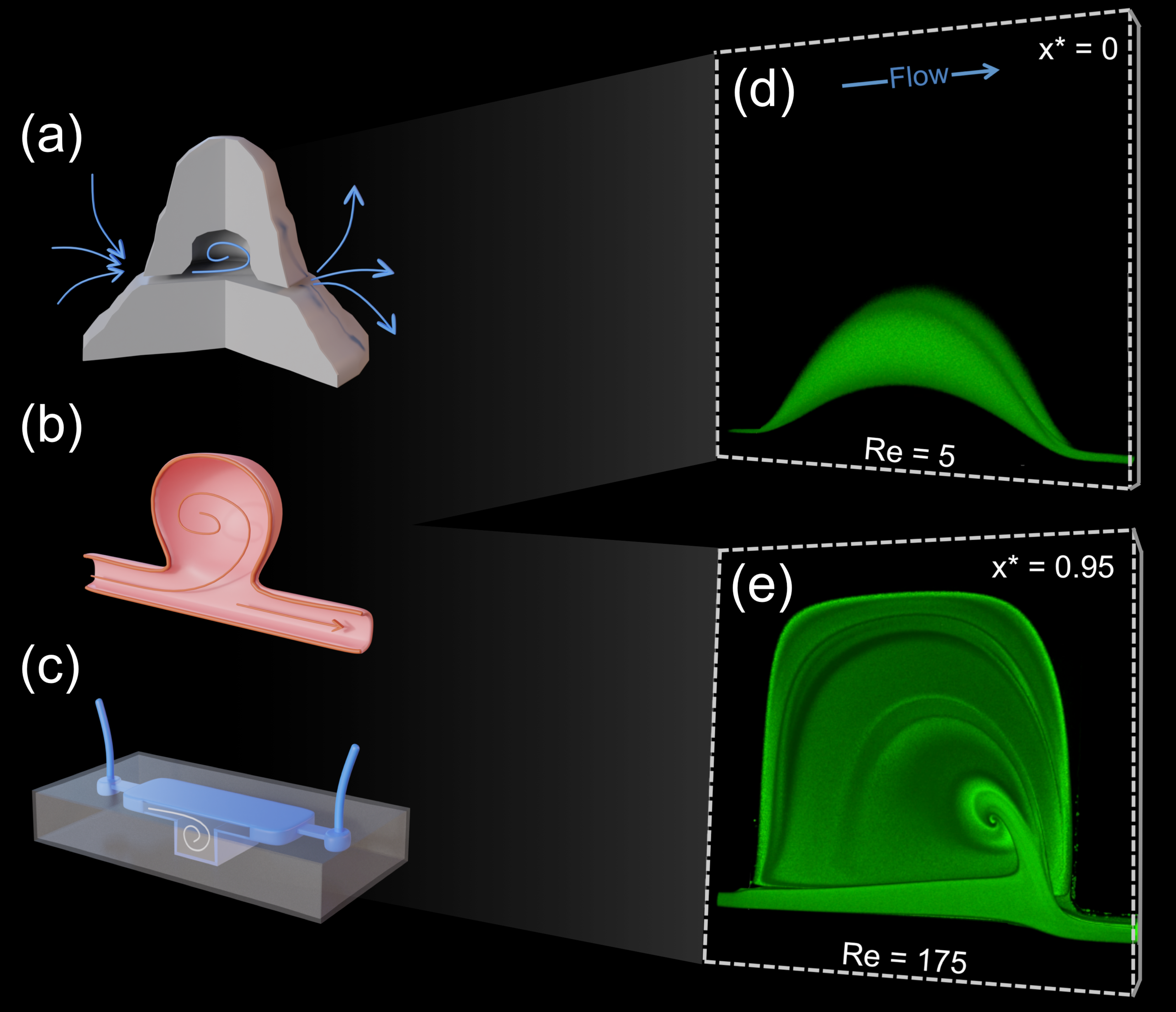}
  \label{fig:1}
  \caption{ Fluid mass transport in three-dimensional cavity flow geometries commonly found in (a) porous ocean rocks, (b) physiological blood flow in aneurysms, and (c) selective cell trapping in microchambers. Confocal 3D imaging of fluorophore containing streams shows flow enters and exits a cavity without circulating at low Reynolds numbers ($Re$) in (d) and fully-developed circulating cavity flow at higher $Re$ in (e) reveals a set of streamlines swirling out of the vortex core and returing to the main channel flow (Fig. S7, video S2).
}
\end{figure}
	
\begin{figure*}
\centering
  \includegraphics[scale=0.159] {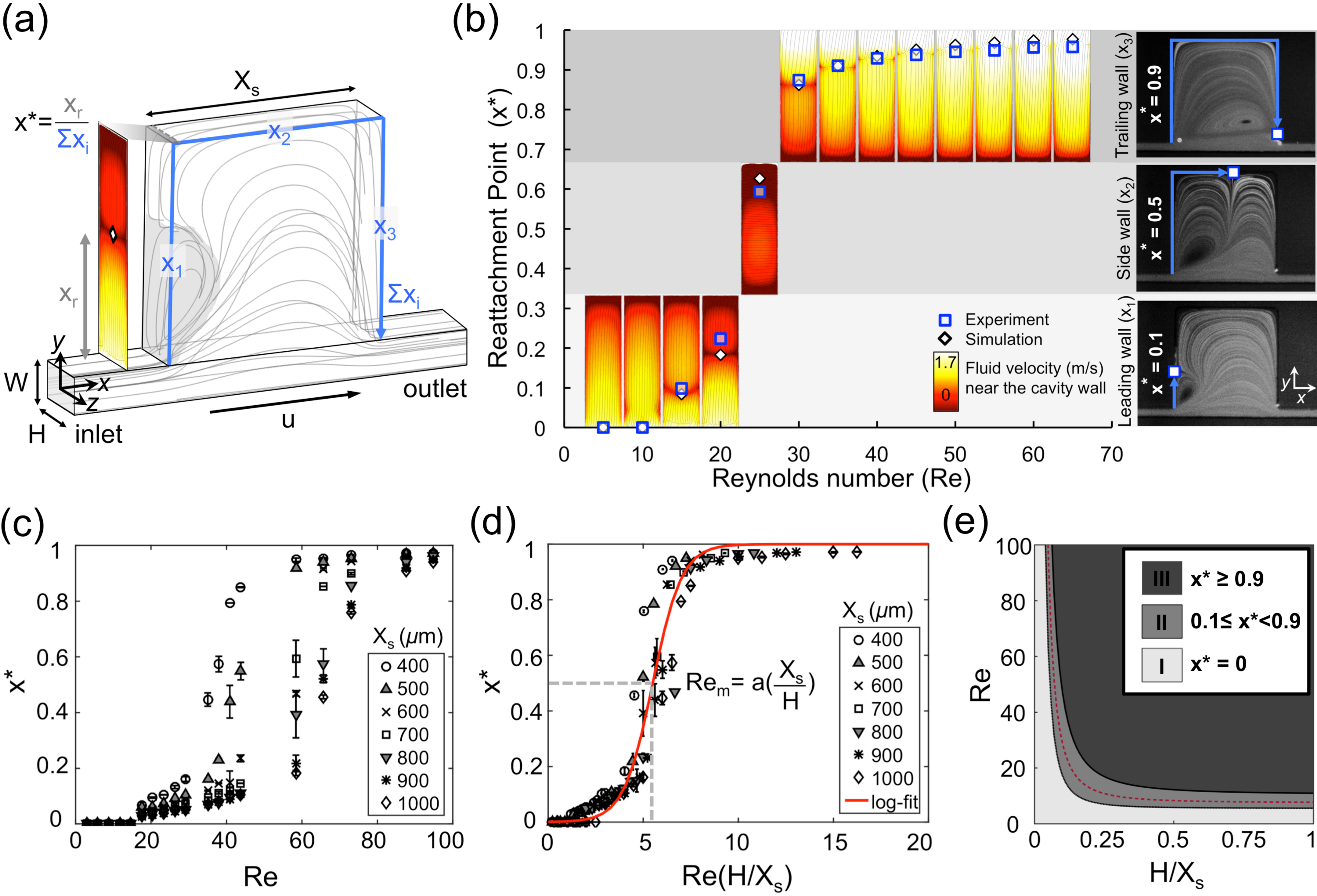}
  \label{fig:2}
  \caption{ Analysis of recirculating vortex flow development in three-dimensional confined cavities (a) Three dimensional flow diagram in a square cavity where the leading wall ($x_{1}$), side wall ($x_{2}$), and lagging wall ($x_{3}$) are equal in length. The formation and growth of recirculating flow in the vortex or wake bubble (gray area) is characterized by the transition of the reattachment point ($x_{r}$), normalized by the length of all stationary cavity walls ($x^{*} = \frac{x_{r}}{\sum x_{i}}$). (b) Experimental ($\square$) and numerical simulation ($\diamond$) of the reattachment point ($x^{*}$) logistic growth and transition on the cavity walls as a function of the channel's Reynolds number ($Re$). (c) Cavity flow development at a range of cavity side lengths collapse in a logistic function presented in a solid red line (-) in (d) when $Re$ is rescaled by the cavity geometry ratio ($\frac{H}{X_{s}}$). (e) Phase diagram of cavity flow development states showing regions where phase I: no circulating cavity flow, phase II: partial cavity flow, and phase III: full cavity flow, as a function of the cavity aspect ratio ($\frac{H}{X_{s}}$) versus the channel Re. The dashed red line is a power-law fit of $Re_{m} \sim (\frac{H}{X_{s}})^{-2}$.
  }
\end{figure*}

There are similarities between the initial stages of vortical flow formation in a confined 3D cavity and the flow through a sudden expansion or backward-facing step at low channel Reynolds numbers \cite{Biswas2004}. Flow separates at the sudden expansion region forming an internal recirculating vortical flow at the leading wall. However, without the existence of a trailing wall to confine the recirculating flow, secondary and tertiary separation bubbles occur downstream in the later stages of such flows \cite{durst1993plane, armaly1983experimental}. In the presence of the trailing wall, flow separation forms a recirculating vortex that grows and occupies the entire cavity.

Although 2D cavity flow has been extensively studied, many aspects of recirculating flow formation and mass transport remains unknown in 3D confined cavities. For example, a defining feature in incompressible and unconfined 2D cavity flow is the formation of a material flux boundary, or separatrix, between the main channel flow and open cavity flow \cite{maull1963three, o1972closed, ghia1982high, horner2002transport}. In contrast, for confined 3D conditions, recirculating inertial flow is highly complex and unfolds in a qualitatively different manner in three-dimensional space \cite{koseff1990complex}, causing unexpected complex flow patterns like the ones generated in wall-confined microchannels \cite{amini2013engineering} and T-junctions \cite{vigolo2014unexpected}. Therefore, more investigations should address the non-trivial dynamics of mass flux in 3D cavities.

In this letter, we further study the unexpected material flux into and out of the cavity flow inspired by initial 3D experimental measurements of flow streams exiting from the core of a cavity to the mainstream channel flow without a mass flux boundary (Fig. 1(d-e)). We found no evidence of a separatrix forming in 3D cavity flow, matching previous numerical studies indicating that the recirculating wake could not be thought of as containing a separatrix \cite{torczynski1993numerical,haddadi2017inertial}. We further show that the streamline linkage, i.e. locations of entry and exit, to the main flow is strongly dependent on the development of the cavity flow. First, we introduced a dimensionless parameter to generalize flow development stages for different cavity geometries. Then we investigated the mass flux through the recirculating cavity flow based on the cavity flow development parameter. Our investigation uncovered an unexpected evolution in the nature of the channel mass flux to the cavity flow as it develops which has implications for the ability to transport particles and cells into cavities.

\subsection{Recirculating cavity flow formation}
Experimental observations in a channel-cavity system with fixed channel flow conditions ($Re  =  u D\textsubscript h/ \nu$), where $u$ is the average inlet flow velocity, $D\textsubscript h$, hydraulic diameter of the channel, and $\nu$ is the kinematic viscosity of the fluid, demonstrate the development of recirculating flow in a square cavity as a function of the cavity height ($H$) and the cavity side length ($X_s$) (Fig. S1 and video S1). Given the complexity in geometry, a Reynolds number alone is not sufficient to define or generalize cavity flow conditions nor predict recirculating flow formation and development in microcavities. One can use a Reynolds number along with a number of non-dimensional geometric parameters instead. Alternatively, one common defining characteristic shared by all cavity geometries is a transition in the location of the reattachment line along the cavity wall at which streamlines diverge between circulating in the cavity, and flowing into the cavity but returning to the main channel without circulating (Fig. 2(a)) \cite{cherry1984unsteady}. The reattachment point $x_r$ as observed from a top-down perspective in the x-y plane, is used to measure the development of recirculating flow in cavities (Fig. S8). To generalize the transition of the reattachment point $x_r$ in different cavity geometries, (${\textstyle x^{*}}{\textstyle=}\frac {x_{r}}{\sum_{}x_i} $) is used as a measure of the reattachment point transition distance along the cavity walls normalized to the total wall lengths of a square cavity with side length $X_s$ (scaling method in supplementary material).

\begin{figure*}
\centering
  \includegraphics[scale=0.149]{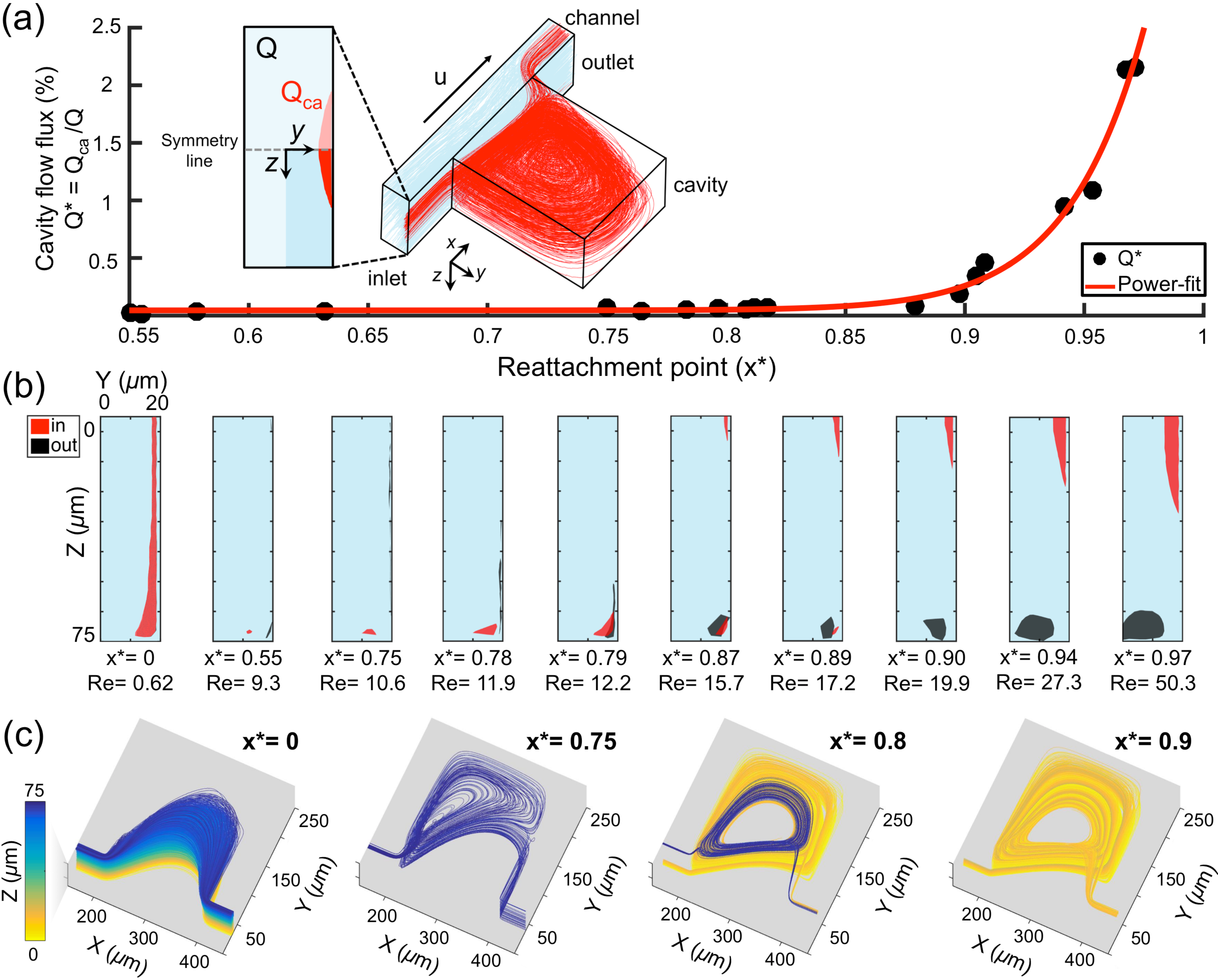}
  \label{fig:3}
  \caption{ Cavity-channel mass flux dynamics as a function of cavity vortex flow development. (a) An exponential increase of cavity flow mass flux $Q^{*}= \frac{Cavity  flux (Q_{ca})}{Channel  flux (Q)}$ is observed when the vortex occupies cavity volume at ($x^{*}>0.9$) and does not significantly expand further in size. The red line is the power-fit law of $Q^{*} \sim x^{*30}$. (b) Entry (red) and exit (black) streamline regions into and out of the recirculating flow are measured at a fixed location upstream and downstream of the cavity, respectively. For a full cavity flow, entry and exit regions switch locations from the side to the middle of the channel wall adjacent to the cavity, and vice versa. The size of the entry region expands towards higher velocity regions of the main channel flow. (c) The evolution of a subset of channel streamlines that enter (and exit) the cavity, color mapped by the by the initial z-height in the channel (Fig S10).
  }
\end{figure*}
The evolution of the reattachment point in a microcavity ($X_s =250 \mu m$, $H=70 \mu m$) varies with the entry channel's Reynolds number ($Re = 0.1-100$) (Fig. 2(b)). At a low Reynolds number, fluid flow passes with fore-aft symmetry through the cavity, as expected for Stokes flow, with no recirculation ($x^*$ = 0) (fig. 1(d)). Flow separates at higher Reynolds number creating a separation bubble with a circulating wake. The vortical flow first evolves over an order of magnitude in Reynolds number, with the reattachment point $x^*$ remaining on the cavity-leading wall ($x_1$) as the Reynolds number increases. Then the reattachment point rapidly migrates along the cavity sidewall ($x_2$), over a small range of Reynolds numbers. Lastly, recirculating flow separation in the cavity reaches an asymptotic phase where the reattachment point stagnates at the end of the cavity trailing wall ($x_3$) ($x^{*}$ $\approx$ 1). The evolution of the reattachment point as a function of the Reynolds number can be described by a logistic function:
  
\begin{equation}
x^{\ast} =\frac{1}{1+e^{-k(Re-a)}} 
\label{eq:2-2}
\end{equation}

in which $x^{*}$ is the estimated reattachment point location, $k$ is the curve steepness, and $a$ is the value of the sigmoid's midpoint. Both simulation and experimental results reflect the logistic function behavior (Fig. 2(b)).

The evolution of the reattachment point follows similar forms for different cavity side lengths ($X_s = 400-1000 \mu m$) (Fig. 2(c)). The vortex growth curves for different cavity side lengths collapse into (Eq. 1) logistic relation (red-line in Fig. 2(d)) when it is rescaled by $Re(\frac{H}X_{s})$, where $k=1.31$ and $a =5.5$ (more data in supplementary material Fig. S2-4). Using this scaling and the logistic function, one can define a condition at $Re = a$ when the vortical flow occupies approximately half of the total cavity volume and the reattachment point reaches the middle of the cavity, i.e. $Re_{m}=Re_{(x^{*}=0.5)} $. $Re_{m}$ is substituted in terms of the logistic function (Eq. 1) to derive a universal expression that estimates full cavity flow at $2Re_{m}$ and no cavity circulating flow at $Re_{m}/2$ for any cavity side length, height, and channel Reynolds number.

\begin{equation}
Re_{m\;}=\;a(\frac{X_{s}}H) \label{eq:2-3}
\end{equation}

To generalize cavity flow transition behavior at any channel-cavity geometry, we studied cavity flow development as a function of the channel $Re$ along with the cavity non-dimensional geometric parameter $H/X_s$. Experimental observations are summarized in a phase diagram (Fig. 2(e)) showing the main cavity flow developmental phases on a range of cavity aspect ratios $0.1\leq H/X_s \leq 1$. Phase I: No cavity flow ($x^{*}=0$) with no recirculation in the cavity. Phase II: partial cavity flow ($0.1 \leq x^{*} < 0.9$) after the formation of a recirculating wake.  Phase III: full cavity flow ($x^{*} \geq 0.9$) when the recirculating vortex fills the cavity volume. The red dotted line represents the power-law fit of $Re_m$ at $x^{*}=0.5$.  The phase diagram suggests an inverse relation between recirculating flow development and cavity aspect ratio. Cavities with low aspect ratio reach full recirculating cavity flow (phase III) at low channel inertial conditions and vice versa. Thus, channel-cavity geometry and flow rate can be designed in a way to either prevent or accommodate recirculating flow.

\subsection{Mass transport in cavity flow}
Given the stages in the development of a recirculating vortex, we asked whether the mass transfer between the main channel and the cavity is dependent on the vortex reattachment point. Three-dimensional finite element method simulations of the incompressible Navier-Stokes equations were used to identify streamlines from the main channel that contribute to the recirculating cavity flow. We plot in Fig. 3(a) the recirculating volumetric flux ($Q^{*}$) in the channel versus the reattachment point ($x^{*}= 0-0.98$) in the cavity. The recirculating flux ($Q^{*}$ = $uA$), where $u$ is the average velocity in $A$ the cross-sectional area, is normalized by the whole channel flux ($Q$) (Fig. S9). Therefore, this represents the percentage of the main channel flow that enters (and exits, -conserving mass) the cavity. Interestingly, the fraction of the main channel flow entering (and exiting) gradually increases with vortex size in the cavity. In particular, an unexpected exponential growth of the channel mass flux contribution to cavity flow is observed when the vortex asymptotically approaches its maximum size after reaching the cavity volume ($x^{*} \approx1$).

Coincident with the increase in mass transport, we observe an unexpected switch in the location of entering and exiting streamlines into the recirculating cavity flow.  Leveraging the flow symmetry in the channel cross-section, we tracked the evolution of the entrance (red) and exit (black) locations across a quarter of the channel (Fig. 3(b)). Before the cavity completely fills ($0.1 \leq x^{*} < 0.9$), fluid enters the vortex from the regions near the upper and lower walls and exit along the middle of the channel side wall downstream of the cavity. Surprisingly, a new entry region grows at the middle of the cavity as the re-attachment point increases eventually leading to the entry and exit regions switching locations as the recirculating flow fills the cavity space further ($x^{*} =0.8-0.9$) (Fig. 3(c) and Fig. S10). At full cavity flow ($x^{*}>0.9$) the topology of the entering streamlines switches completely such that streamlines enter at the middle of the channel sidewall. Coincident with the mass flux increasing, the entry region area grows towards the center of the channel as the re-attachment point and Reynolds number increases further. This increase in the entry area spanning high-velocity regions of the channel cross-section explains the exponential increase of the recirculating flux in the cavity.

Confocal imaging experiments of selectively dyed streams also reflect the mass flux into and out of the recirculating flow in the cavity. In our experimental set-up, we increment the location in the channel cross-section of fluorescently labeled streams by controlling flow rates: Q1 (rhodamine: red dye), Q2 center (fluorescein: green dye), and Q3 (water: colorless), and with Q1=Q2=Q3 set at the same flow rate (Fig. S5). A significant decrease in the contribution of Q2 streamlines in the recirculating flow is observed in the transition between ($x^{*}=0.8-0.9$) over which flow entry streamlines are modeled to change location from the center to the sidewall. For a full cavity flow ($x^{*}>0.9$), we visualize in 3D the flow exit from the cavity at the top and bottom of the channel leaving from the core of the vortex. Here streamlines are selectively stained as follows: Q2 (fluorescein: green dye), Q1, and Q3(water: colorless) (Fig.1(e) and Fig. S6 and video S2). In a new cavity design with a notch or step partially blocking the cavity outlet region, fluorescently labeled streamlines are observed leaving from the core while diverting from the notch and returning to the mainstream channel flow. Thus, the flow exit behavior remained consistent with geometric changes at the cavity outlet region (Fig. S7).

Our numerical and experimental results in sum indicate that there is direct mass transport from the channel flow into the recirculating flow in the cavity with no mass flux boundary or separatrix. Besides showing that streamlines from the main flow enter the recirculating flow and later leave, we found no evidence of closed streamlines in the recirculating flow that are not connected to the main flow. Previous theoretical investigation suggests the same manner of continuous mass exchange between the channel and the cavity when a full cavity flow is reached \unskip~\cite{cherry1984unsteady}. Our results shed new light on the increase in the relative mass flux (as a function of main channel flow rate) as the cavity flow reaches a fully-filled configuration, as well as the shifting locations of influx and efflux from the cavity. The enhanced transport between the cavity and the channel as well as shifting streamline locations that leave the cavity can contribute to the depletion of particles captured in microvortices at later stages of cavity flow as observed in previous studies \unskip~\cite{khojah2017size}. This change in fluid flux direction and magnitude can modulate the limit cycle trajectory of trapped particles, also experiencing inertial lift forces, to exit the vortex as the recirculating flow envelops the entire cavity  \unskip~\cite{haddadi2017inertial}.
	
These findings expand our understanding of mass transfer in 3D cavity flow with finite inertia -a widely observed geometry across nature and engineered systems.  Our findings can inform the design of cavities to avoid dead zones of low mass transport, or enhance transport from specific entry locations in the main flow to e.g. isolate particles. The understanding of how mass transport is not limited across a separatrix also has implications for the evolution of biochemical processes in microvortices inside ocean rocks, physiological flow and micro-clot accumulation in aneurysms, and engineering microwells for cell trapping in miniaturized diagnostic devices.

\section*{Acknowledgements}
We Thank Dr. Dan Stoecklein and Dr. Kaytlien Hood for helpful discussion and input. We gratefully acknowledge the support of the Advanced Light Microscopy/Spectroscopy Center (ALMS) at the California NanoSystems Institute (CNSI) in the University of California, Los Angeles.

\nocite{*}

\bibliography {aipsamp}   

\end{document}